\documentclass[final]{elsart3p}

\journal{J. Crystal Growth}
\date{12 February 2008}

\usepackage{graphicx}
\usepackage{textcomp}
\usepackage{amssymb}
\usepackage{amsmath}

\begin{document}

\begin{frontmatter}

\title{The growth of ZnO crystals from the melt}

\author{D. Klimm\corauthref{cor1}}\ead{klimm@ikz-berlin.de},
\author{S. Ganschow},
\author{D. Schulz}, and
\author{R. Fornari}

\corauth[cor1]{corresponding author}

\address{Institute for Crystal Growth, Max-Born-Str. 2, 12489 Berlin, Germany}

\begin{abstract}
The peculiar properties of zinc oxide (ZnO) make this material interesting for very different applications like light emitting diodes, lasers, and piezoelectric transducers. Most of these applications are based on epitaxial ZnO layers grown on suitable substrates, preferably bulk ZnO. Unfortunately the thermochemical properties of ZnO make the growth of single crystals difficult: the triple point 1975$^{\,\circ}$C, 1.06\,bar and the high oxygen fugacity at the melting point $p_{\mathrm{O}_2} = 0.35$\,bar lead to the prevailing opinion that ZnO crystals for technical applications can only be grown either by a hydrothermal method or from ``cold crucibles" of solid ZnO. Both methods are known to have significant drawbacks. Our thermodynamic calculations and crystal growth experiments show, that in contrast to widely accepted assumptions, ZnO can be molten in metallic crucibles, if an atmosphere with ``self adjusting" $p_{\mathrm{O}_2}$ is used. This new result is believed to offer new perspectives for ZnO crystal growth by established standard techniques like the Bridgman method.
\end{abstract}


\end{frontmatter}

\section{Introduction}

The wide band gap (3.3\,eV, direct transition) and the large exciton binding energy (60\,meV) make zinc oxide (ZnO) an interesting material for optoelectronic applications. ZnO crystallizes in the wurtzite structure and is therefore piezoelectric. Its piezoelectricity and the dependence of the electric resistance on environmental influences can be used for sensors. However, the broad application of ZnO single crystals is still hindered by the insufficient supply of cheap and good substrates for the production of high quality epitaxial layers. ZnO could be a perfect material for the production of wafers, but unfortunately the bulk growth of ZnO crystals is difficult. Like most other semiconductor compounds (e.g. GaAs, GaP) ZnO melts under decomposition and the volatile component O$_2$ evaporates. In the case of GaAs (melting point $T_\mathrm{f}=1238^{\,\circ}$C, total vapor pressure at $T_\mathrm{f}$: $p_\mathrm{tot} = 1-3$\,bar) and GaP ($T_\mathrm{f}=1467^{\,\circ}$C, $p_\mathrm{tot} = 35\pm10$\,bar) \cite{Richman63,Wenzl93} the evaporation of volatile As$_4$ or P$_4$ can be suppressed by an ambient pressure in the growth chamber that is slightly larger than $p_\mathrm{tot}$. Fortunately, a variety of materials is available that withstands the thermal and chemical conditions during crystal growth of GaAs and GaP. B$_2$O$_3$ can be used as ``liquid encapsulant'' and fused silica (SiO$_2$), pyrolytic boron nitride (BN) or graphite (C) are suitable for crucibles, heaters and other constructive parts of the crystal growth set-up.

The situation is more complicated for zinc oxide. The triple point of this substance is at $T_\mathrm{f}=1975^{\,\circ}$C and $p_\mathrm{tot} = 1.06$\,bar. To maintain a ZnO melt stable the total pressure must be considerably larger than 1.06\,bar. At the melting point the substance evaporates under dissociation
\begin{equation}
\mathrm{ZnO} \rightleftarrows \mathrm{Zn} + \frac{1}{2}\, \mathrm{O}_2
\label{eq:ZnO}
\end{equation}
resulting in an equilibrium oxygen fugacity $p_{\mathrm{O}_2} = 0.35$\,bar \cite{FactSage5_5}. The choice of materials (especially for crucibles) that withstand temperatures close to $2000^{\,\circ}$C under such highly oxidative conditions is very limited: The electrochemical potential of the reaction Me + O$_2$ $\longrightarrow$ MeO$_2$ at $1975^{\,\circ}$C and 20\,bar is negative only for Me = iridium and osmium ($E=-0.1474$\,V for Ir and $E=-0.0878$\,V for Os). It must be noted that even Ir is considerably less noble than platinum ($E=-0.4422$\,V), but unfortunately Pt melts already at $1768^{\,\circ}$C. Generally it is assumed that Ir crucibles cannot be used for the melt growth of ZnO crystals as the metal would quickly be oxidized \cite{Nause99}. If one works under less oxidative conditions, the equilibrium reaction (\ref{eq:ZnO}) may lead to the formation of metallic zinc that alloys and finally destroys Ir crucibles.

As a result of these problems, technical applications of ZnO (varistors, sensors, transparent conducting oxide) are mainly based on powders, on polycrystalline or epitaxial layers, or on ceramics. The growth of ZnO single crystals is possible from oxide fluxes \cite{Wanklyn70}, by sublimation \cite{Helbig72,Look98}, or from hydrothermal solutions \cite{Ohshima04}. Nowadays, the hydrothermal method allows to grow the best ZnO bulk crystals that are available and wafers up to 2 inch diameter can be produced \cite{Maeda05}. However, crystals grown from oxide or hydrothermal fluxes usually contain traces of the solvent and the rate of sublimation growth (that works without flux) is low.

A high pressure technique for the bulk ZnO growth from cold crucibles is proposed to solve these problems \cite{Nause99,Nause05,Reynolds04}. Here an \emph{rf} field is coupled directly to the ZnO melt. Already in 1941 Miller found by measurements that the electrical conductivity $\sigma_\mathrm{el}$ of ZnO rises with $T$ at least up to $700-800^{\,\circ}$C \cite{Miller41}. The same positive correlation $\sigma_\mathrm{el}(T)$ was observed for melts of other oxides up to $1630^{\,\circ}$C and is a common characteristic of ionic liquids \cite{Schiefelbein97}. Hence it must be assumed that thermal fluctuations of oxide melts used in a cold crucible process are amplifying, as hot spots within the melt are heated more by the \emph{rf} power than colder spots. As a result, such ``skull melting" methods work with very large temperature gradients impairing crystals quality.

In this article it will be shown that the growth of ZnO single crystals is possible by an almost classical technology. ZnO melt can be held stable and can be crystallized in iridium crucibles, if a suitable atmosphere is used during the growth process.

\section{Thermodynamics}

\begin{figure*}[htb]
\includegraphics[width=0.60\textwidth]{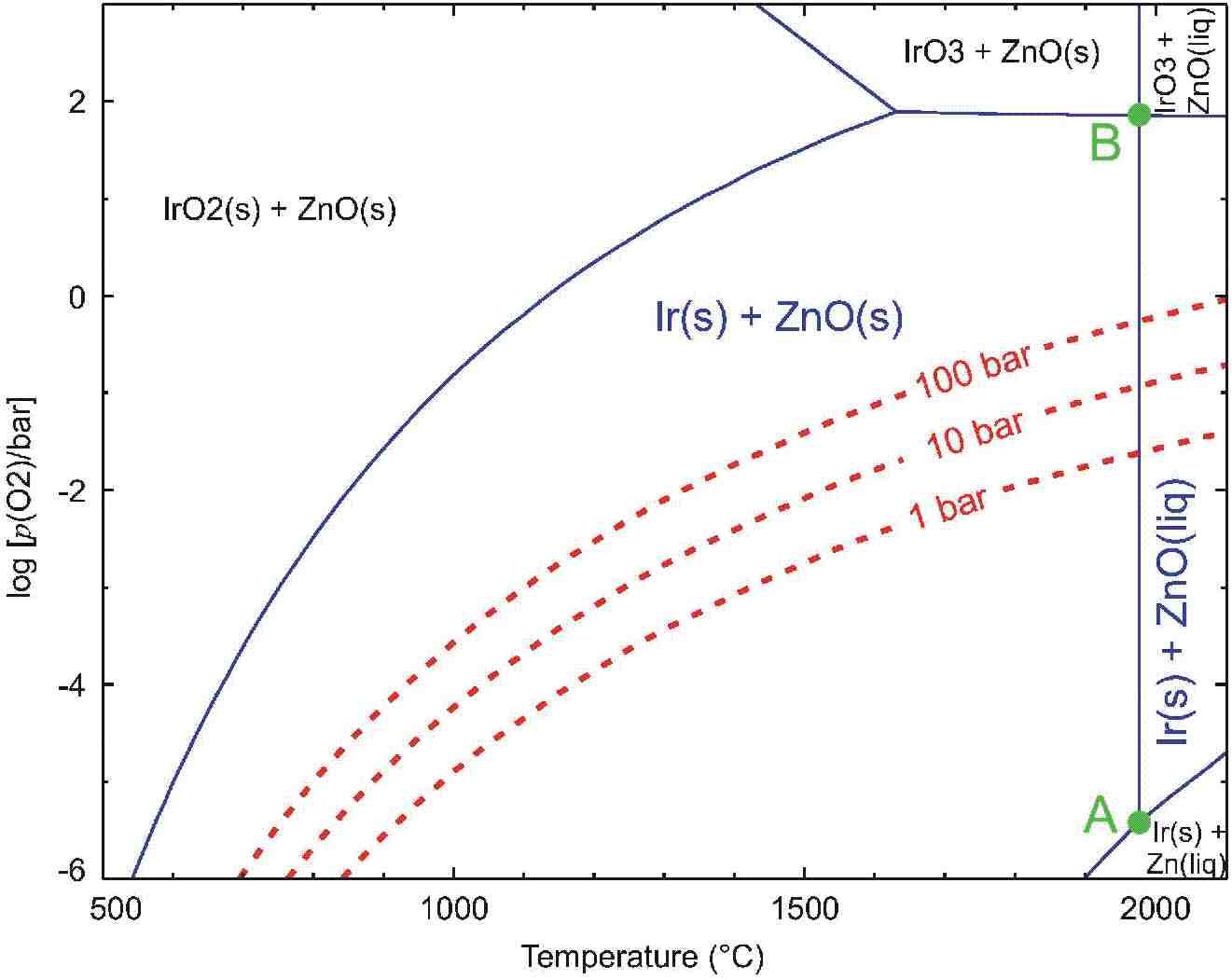}
\caption{\label{fig:Ir-Zn-O2} Predominance diagram Zn--Ir--O$_2$ showing stability regions for metals and oxides in dependence on $T$ and the oxygen fugacity $p_{\mathrm{O}_2}$. The dashed lines are the equilibrium $p_{\mathrm{O}_2}$ of pure CO$_2$ for total pressures of 1, 10, and 100\,bar, respectively.}
\end{figure*}

A quantitative thermodynamic treatment of the system Ir--Zn--O$_2$ with FactSage \cite{FactSage5_5} shows that, despite the claims given by Nause \cite{Nause99}, the growth of bulk ZnO from the melt should be possible using Ir crucibles. The predominance diagram (solid lines in Fig.~\ref{fig:Ir-Zn-O2}) shows that at $T_\mathrm{f}=1975^{\,\circ}$C solid and liquid ZnO is in equilibrium with Ir metal between points \sffamily A\normalfont\ and \sffamily B\normalfont\ for $-5.431\leq\log\left[p_{\mathrm{O}_2}/\mathrm{bar}\right]\leq+1.831$. Along this melting isotherm of ZnO $p_\mathrm{Zn}$ drops from 214\,bar at the triple point with Zn(liq) (\sffamily A\normalfont) to 0.051\,bar at point \sffamily B\normalfont\ where Ir disappears. In the central part of the melt isotherm one calculates $p_\mathrm{Zn}=4.19$\,bar for $p_{\mathrm{O}_2}=0.01$\,bar and $p_\mathrm{Zn}=1.32$\,bar for $p_{\mathrm{O}_2}=0.1$\,bar. From (\ref{eq:ZnO}) follows, that $p_\mathrm{Zn}$ drops by one order of magnitude if $p_{\mathrm{O}_2}$ is lowered by two orders of magnitude.

Fig.~\ref{fig:Ir-Zn-O2} shows, however, that it is impossible to find a mixture of O$_2$ with an inert gas like argon that is suitable to keep Ir and ZnO chemically stable for the whole crystal growth process, as all $p_{\mathrm{O}_2}=$ const. are represented by horizontal lines in the diagram. Any gas containing a constant O$_2$ admixture crosses somewhere the {IrO$_2$(s)+ZnO(s)}/{Ir(s)+ZnO(s)} pha\-se boundary. Indeed it is well known that an atmosphere where Ir devices are to be heated must not contain more than 1--2\% O$_2$ to avoid burning of the metal. One could try to heat the growth set-up in an oxygen free atmosphere to $1000-1200^{\,\circ}$C and to add O$_2$ later, but practically this is not a solution for the problem: Every crystal growth process from the melt has to maintain $T$ gradients to establish stable growth. Within such gradients some Ir parts would always be at such low $T$ that the Ir stability region is left and IrO$_2$ is formed. Similar problems during the melt growth of $\beta$-Ga$_2$O$_3$ were solved recently by working in an Ar/CO$_2$ gas mixture \cite{Tomm00}. The dissociation reaction
\begin{equation}
\mathrm{CO}_2 \rightleftarrows \mathrm{CO} + \frac{1}{2}\, \mathrm{O}_2
\label{eq:CO2}
\end{equation}
results in an oxygen partial pressure that depends on $T$ and $p_\mathrm{tot}$. The $p_{\mathrm{O}_2}$ that results from (\ref{eq:CO2}) for $p_\mathrm{tot}=1,10$, and 100\,bar are indicated by dashed lines in Fig.~\ref{fig:Ir-Zn-O2}. It turns out, that the oxygen amount that is supplied by (\ref{eq:CO2}) depends similarly on $T$ like the IrO$_2$/Ir and the ZnO/Zn phase boundaries. Reaction (\ref{eq:CO2}) at elevated $T$ proceeds much faster than the flow of gas. This means that every part (Ir crucible, insulation materials, ZnO sample) within the growth chamber is surrounded by a $p_{\mathrm{O}_2}(T)$ that depends on $T$ in an advantageous way --- $p_{\mathrm{O}_2}(T)$ is ``self adjusting".

\section{Crystal growth}

\begin{figure}[htb]
\includegraphics[width=0.35\textwidth]{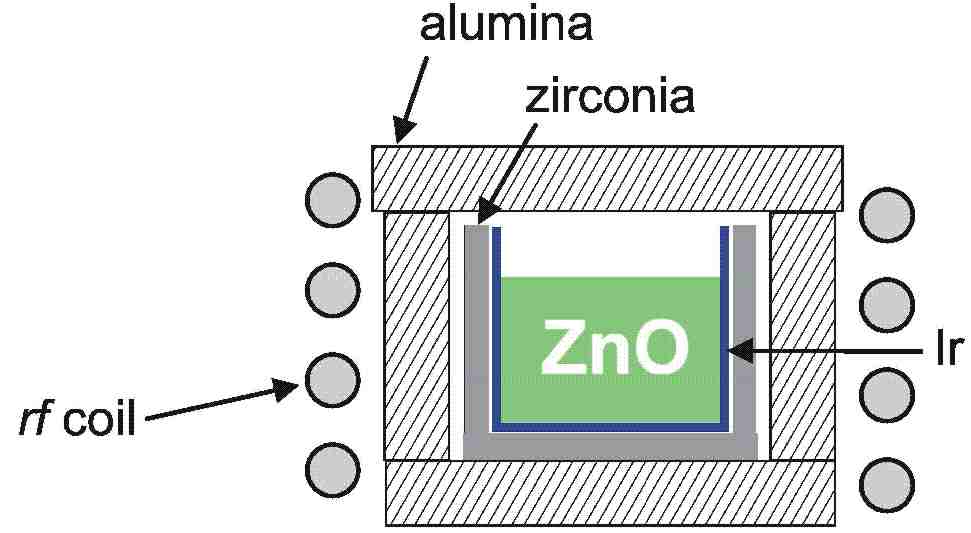}
\caption{\label{fig:melt} ZnO melt set-up with \emph{rf} heating and ZrO$_2$ + Al$_2$O$_3$ ceramics as thermal insulation. The whole set-up is placed inside a pressure vessel that is not shown here.}
\end{figure}

First melting experiments were carried out in a high-pressure Czochralski apparatus with inductive heating. ZnO powder (Alfa Aesar, 99.99\% purity) was pressed isostatically to reduce its volume and then loaded to a 40\,ml iridium crucible. The crucible was surrounded by alumina and zirconia insulating ceramics and placed inside the induction coil (Fig. 2). The growth chamber was evacuated and subsequently filled with (a) a mixture of 6\,vol-\% CO$_2$ in Ar, and (b) with pure CO$_2$ to a pressure of 17.5\,bar. The crucible was heated up until the ZnO powder melted. At this point the total pressure inside the chamber reached approximately 19\,bar. After a short time the heating power was switched off and the crucible cooled down to room temperature in about two hours.

\begin{figure}[htb]
\includegraphics[width=0.35\textwidth]{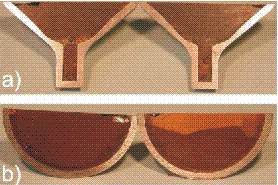}
\caption{\label{fig:Bridgman} Side view (a) and top view (b) of the conical part of a Bridgman crucible with 2 ZnO subgrains (crucible diameter: 38\,mm).}
\end{figure}

After cooling, the solidified melt formed a compact body of dark brown color. No remarkable dependence on atmosphere (a) or (b) could be observed. The largest grains found in this body made it possible to prepare single crystalline blocks with up to 5\,mm edges oriented along the crystal's principal axes. The possibility to hold a stable ZnO melt within metallic crucibles without significant corrosion of the crucible is an important precondition for the creation of a melt-based growth process \cite{Bertram04c}. Such processes like the Czochralski- or Bridgman-method should run close to thermodynamic equilibrium with moderate $T$ gradients of some 10\,K\,cm$^{-1}$ or even lower and growth rates of a few millimeter per hour -- in contrast to the skull melting process \cite{Nause99,Nause05} where gradients are at least one order of magnitude larger.

The crystalline quality could be improved using crucibles with conical bottom and a seed channel in a Bridgman-like configuration \cite{Schulz06}. It turned out, however, that good seeding is often prohibited by voids or iridium needles close to the seed/bulk interface. Under unseeded growth conditions, directions almost parallel to the $\vec{c}$ axis were found to be the favorite growth direction, e.g. $[10\bar{1}5]$ which is $12^\circ$ tilted from $[0001]$ \cite{Frank65}. The large grains in Fig.~\ref{fig:Bridgman} a) and b) correspond to $[10\bar{1}5]$ and $[0001]$. If cooling of the crucibles is performed with a rate of $2-10$\,K per hour and if seeding is reached by proper $T$ gradients, solidified ZnO bulks can be produced that are built up from one or a few single crystals. One bulk that was separated from the crucible by a hollow drill is shown in Fig.~\ref{fig:B29}. The following problems remain:

\begin{figure}[htb]
\includegraphics[width=0.20\textwidth]{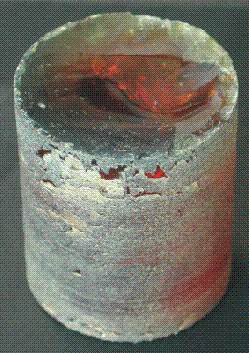}
\caption{\label{fig:B29} Cylindrical part (diameter 33\,mm) of a Bridgman crystal. The bottom as shown in Fig.~\ref{fig:Bridgman} is already cut off.}
\end{figure}

\begin{figure}[htb]
\includegraphics[width=0.45\textwidth]{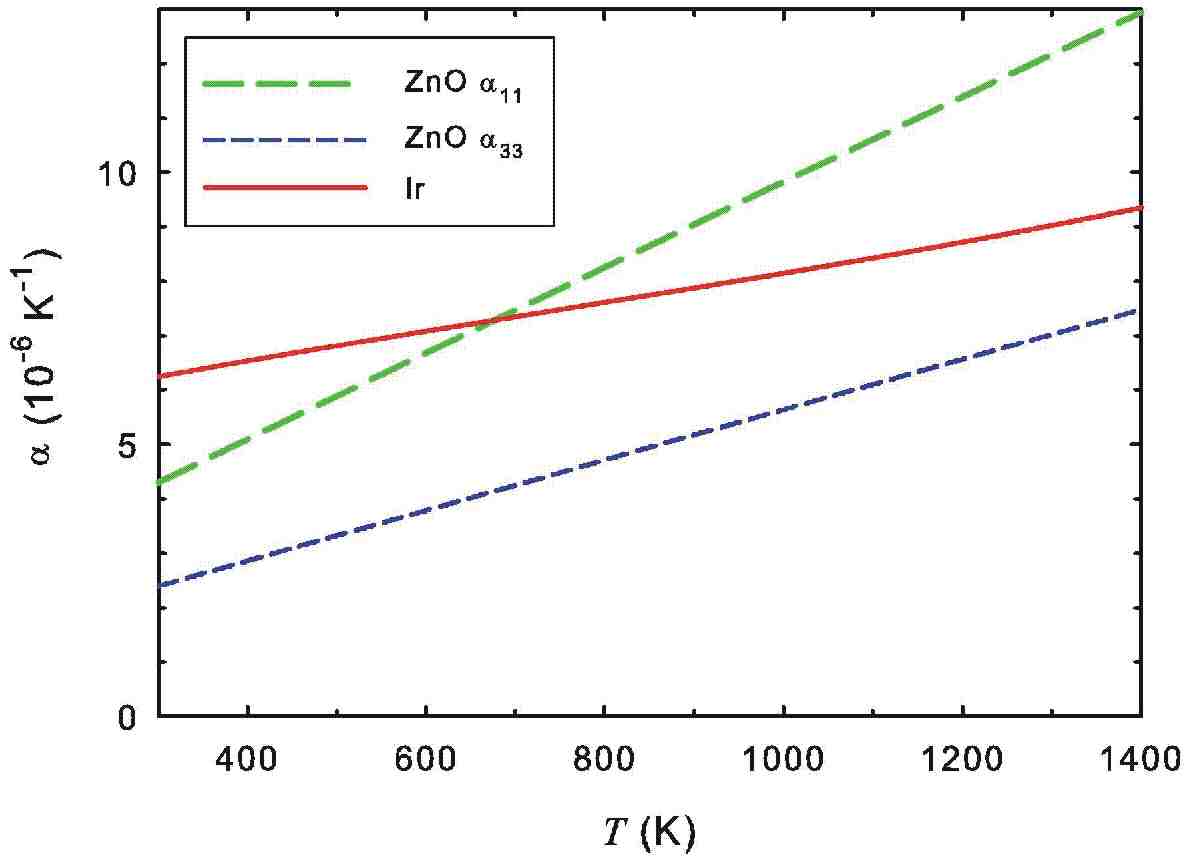}
\caption{Thermal expansion coefficients $\alpha$ for iridium \cite{Halvorson72} and zinc oxide \cite{Iwanaga00}.}
\label{fig:expansion}
\end{figure}

\begin{enumerate}
\item As seen in Fig.~\ref{fig:Bridgman}~(a), dendritic Ir needles with $\leq2$\,mm length grow mainly around the seed region. It is not yet completely clear whether or not these needles disturb the ZnO crystal growth.
\item Most of the crystals contain small angle grain boundaries or are even cracked. One must assume, that the different thermal expansion coefficients of the iridium crucible and the growing crystal are responsible for thermal stresses --- especially, as ZnO was found to stick hard to the crucible wall. The expansion coefficient of the cubic and polycrystalline iridium metal is isotropic, whereas ZnO has parallel to the $\vec{c}$ axis the coefficient $\alpha_{33}$ and perpendicular to $\vec{c}$ $\alpha_{11}$. Fig.~\ref{fig:expansion} compares values for both substances that were measured for $0\leq T \mathrm{(K)}\leq1373$. If the crystal grows along the conventional axis $\vec{c}=[0001]$, $\alpha_{11}(T)$ for ZnO should be compared with $\alpha(T)$ for Ir. The measured data show, that for $T\gtrsim405^{\,\circ}$C (678\,K) $\alpha_{11}$ for ZnO is larger than the expansion coefficient for Ir (Fig.~\ref{fig:expansion}). Consequently, the ZnO bulk crystallizing at $T_\mathrm{f}=1975^{\,\circ}$C is expected to contract during cooling more than the Ir crucible and the cooling process of the $\vec{c}$ oriented crystal within the crucible is not considered to be the reason for the observed small angle boundaries and cracks. Instead, we assume the crystallization process itself as critical: Unfortunately, no reliable data on the mass density of liquid ZnO are available, but it is well known that wurtzite type ZnO (the only solid equilibrium phase under ambient pressure) transforms at $9.1\pm0.2$\,GPa to the rocksalt structure under as much as 16.7\% volume collapse \cite{Bates62,Desgreniers98}. Obviously, wurtzite type ZnO (coordination number 4) is comparably loose packed whereas the mass density $\varrho$ of rocksalt type ZnO (coordination number 6) is higher. If one assumes that liquid ZnO is approximately close packed as the rocksalt phase, a substancial expansion must be expected upon crystallization of ZnO from the melt. Analogous drops of $\varrho$ by $\approx9$\% upon crystallization were reported for other tetrahedrally coordinated semiconductors like Si \cite{Logan59} or GaAs \cite{Jordan80}, respectively.
\end{enumerate}

\begin{figure}[htb]
\includegraphics[width=0.46\textwidth]{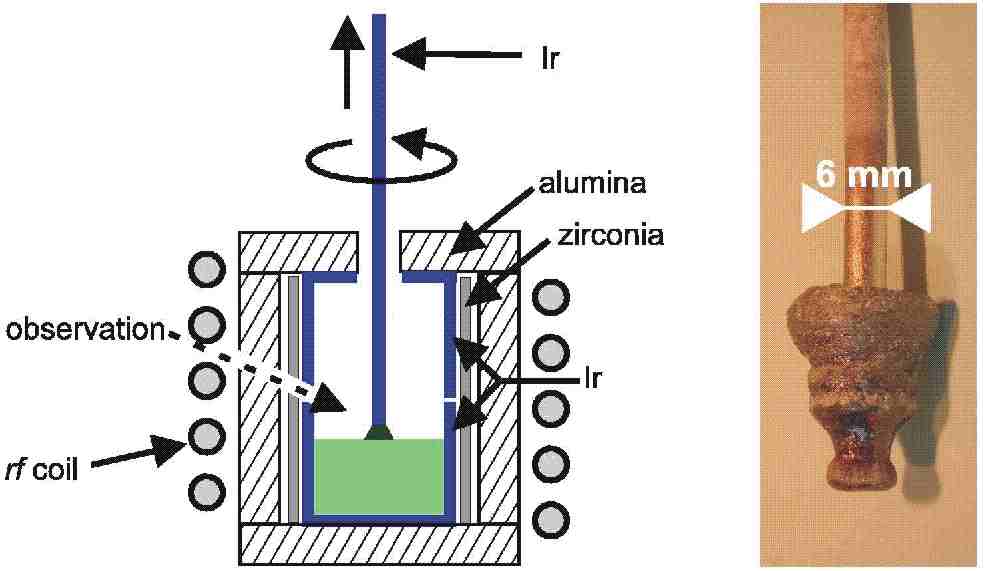}
\caption{Left: Czochralski set-up used for ZnO melts. Right: Polycrystalline ZnO solidified on the Ir seed rod.}
\label{fig:Czochralski}
\end{figure}

To avoid cracking issues as mentioned above, a series of Czochralski growth experiments with ZnO was performed. Unfortunately, all attempts to obtain crystallization of bulk ZnO failed. Only polycrystalline ZnO aggregated on the iridium rod that was used instead of a seed and the solidification process stopped after $\lesssim20$\,mm (Fig.~\ref{fig:Czochralski}). The control of the Czochralski experiments failed as the evaporation rate of ZnO at $T_\mathrm{f}$ is high, even under the pressures up to 20\,bar that were used. The evaporated material forms white fume laying above the melt's surface making optical control of the seeding process almost impossible. Moreover, evaporated ZnO condenses partially on the Ir seed rod, especially where it is lead through the thermal insulation. The sublimate creates mechanical contact between both parts, thus hindering mass control of the crystallizing ZnO by the balance on top of the seed rod that would otherwise allow automatic diameter control of the Czochralski growth process.

Generally, the Czochralski process is difficult to control at very high temperatures $T_\mathrm{f}\gtrsim1800^{\,\circ}$C, as a sufficient heat flow density $P$ resulting in thermal gradients $\nabla T$ of several 10\,K\,cm$^{-1}$ must be maintained from the hot melt/crystal interface ($T_1=T_\mathrm{f}$) to some heat sink at the top of the crystal ($T_2<T_1$) for the performance of a stable growth process. $P$ consists of a conductive part $P_\mathrm{cond}$ and a radiative part $P_\mathrm{rad}$

\begin{eqnarray}
P               & = & P_\mathrm{cond} + P_\mathrm{rad} \quad\mathrm{with}\\
P_\mathrm{cond} & = & \lambda(T) \nabla T \label{eq:cond} \\
P_\mathrm{rad}  & = & \sigma (T_1^4 - T_2^4) \label{eq:rad}
\end{eqnarray}

where $\sigma=5.67\times10^{-8}$\,W\,m$^{-2}$\,K$^{-4}$ is the Stefan-Boltzmann constant and $\lambda$ is the thermal conductivity. For $T\gg\Theta$ (Debye temperature) $\lambda$ drops often due to increased phonon scattering. For polycrystalline ZnO samples with 1.5\% porosity a drop from 32\,W\,m$^{-1}$\,K$^{-1}$ ($T=300$\,K) to 9\,W\,m$^{-1}$\,K$^{-1}$ ($T=1200$\,K) was measured \cite{Gadzhiev03}, thus reducing $P_\mathrm{cond}$ (\ref{eq:cond}). Usually $P_\mathrm{rad}$ rises drastically with $T$ if the optical transparency in the relevant near infrared spectral region is sufficiantly high (\ref{eq:rad}). For materials with bad transparency the contribution of $P_\mathrm{rad}$ cannot replace the smaller contribution of $P_\mathrm{cond}$ and crystal growth becomes unstable, often leading to ``spiral growth''. Such growth instabilities were reported for gadolinium gallium garnet Gd$_3$Ga$_5$O$_{12}$ \cite{Naumowicz81}, rutile TiO$_2$ \cite{Machida94}, and recently for different rare-earth scandates REScO$_3$ \cite{Uecker06b}. The combined low thermal conductivity and opaqueness of ZnO close to $T_\mathrm{f}$ could prevent the Czochralski growth of ZnO too (Fig.~\ref{fig:Czochralski}).

As grown crystals are brownish colored and the intensity of the color depends on cooling rate and $p_{\mathrm{O}_2}$ during growth: Lowering $p_{\mathrm{O}_2}$ and fastening the cooling process increases the intensity of coloring. Recently \cite{Schulz06} it has been pointed out that annealing at $900^{\;\circ}$C bleaches out ZnO crystals almost completely. Already a 1~hour treatment at this temperature is sufficient to shift the absorption edge of our melt grown material from initially $\approx550$\,nm to $\lesssim400$\,nm, whereas the color of ZnO crystals grown by chemical vapour transport changed from reddish orange to transparent within $3-5$~hours at $900^{\;\circ}$C and was accompanied by a change of photoconductivity \cite{Sato06}.

Electrical properties of the current material were investigated by Hall effect measurements. Typical parameters for as grown and nominally undoped Bridgman crystals are: $n$-type conductivity, resistivity 0.15\,$\Omega$cm, electron concentration $2.1\times10^{17}$\,cm$^{-3}$, electron mobility 200\,cm$^2$\,V$^{-1}$\,s$^{^-1}$. The structural quality of the wafers was characterized by high resolution X-ray diffraction. As alrady reported in \cite{Schulz06} down to 22\,arcsec were found for the 0002 reflection.

\section{Discussion and outlook}

Whenever possible, the growth of single crystals on an industrial scale is performed from the melt. The high mass transport rate in melts as compared to solutions or gaseous phases, respectively, is one reason for this choice. Moreover, crystals grown from solutions incorporate usually traces of the solvent which might be a drawback if high purities are required e.g. for electronic applications. The application of carbon dioxide as reactive component in a ``self adjusting atmosphere'' allows the growth of ZnO bulk crystals from iridium crucibles using a conventional Bridgman technology. (Zn,Mg)O alloys were shown to grow pseudomorphically over several 100\,nm thickness on (0001) wafers that were produced from such ZnO boules \cite{Sadofev07}. The undoped ZnO crystals exhibit the typical $n$-type conductivity and a high electron mobility. A further improvement of the electric properties of the material, especially a further reduction of the carrier concentration, is expected feasible with ZnO starting material of higher purity compared with the 4N material that was used for the present investigation. Small angle grain boundaries and cracks are the most important problems to be solved for the development of a melt growth technology for ZnO crystals from iridium crucibles. A further improvement of the electric properties of the material, especially a further reduction of the carrier concentration, is expected feasible with ZnO starting material of higher purity compared with the 4N material that was used for the present investigation.

\ack{

The authors express their gratitude to Karin Struve for help with the Bridgman experiments and to Klaus Irmscher and Martin Schmidbauer for electrical and X-ray measurements. This work was supported by the ``European Regional Development Fund'' (contract 10133344) and by the Leibniz Association (``Pact for Research and Innovation'').}


\begin{thebibliography}{99}

\bibitem{Richman63}
D.~Richman, J. Phys. Chem. Solids 24 (1963) 1131.

\bibitem{Wenzl93}
H.~Wenzl, W.~A. Oates, K.~Mika, in: Handbook of Crystal Growth, Vol.~1A, Elsevier, Amsterdam, 1993, 103.

\bibitem{FactSage5_5}
GTT Technologies, Kaiserstr. 100, 52134 Herzogenrath, Germany, FactSage 5.5, \ttfamily http://www.factsage.com/ \normalfont (2007).

\bibitem{Nause99}
J.~E. Nause, III-Vs Review 12 (1999) 28.

\bibitem{Wanklyn70}
B.~M. Wanklyn, J. Crystal Growth 7 (1970) 107.

\bibitem{Helbig72}
R.~Helbig, J. Crystal Growth 15 (1972) 25.

\bibitem{Look98}
D.~C. Look, D.~C. Reynolds, J.~R. Sizelove, R.~L. Jones, C.~W. Litton, G.~Cantwell, W.~C. Harsch, Solid State Commun. 105 (1998) 399.

\bibitem{Ohshima04}
E.~Ohshima, H.~Ogino, I.~Niikura, K.~Maeda, M.~Sato, M.~Ito, T.~Fukuda, J. Crystal Growth 260 (2004) 166.

\bibitem{Maeda05}
K.~Maeda, M.~Sato, I.~Niikura, T.~Fukuda, Semicond. Sci. Technol. 20 (2005) S49.

\bibitem{Nause05}
J.~Nause, B.~Nemeth, Semicond. Sci. Technol. 20 (2005) S45.

\bibitem{Reynolds04}
D.~C. Reynolds, C.~W. Litton, D.~C. Look, J.~E. Hoelscher, B.~Claflin, T.~C. Collins, J.~Nause, B.~Nemeth, J. Appl. Phys. 95 (2004) 4802.

\bibitem{Miller41}
P.~H. Miller, Phys. Rev. 60 (1941) 890.

\bibitem{Schiefelbein97}
S.~L. Schiefelbein, D.~R. Sadoway, Metallurgical and Materials Transactions B 28 (1997) 1141.

\bibitem{Tomm00}
Y.~Tomm, P.~Reiche, D.~Klimm, T.~Fukuda, J. Crystal Growth 220 (2000) 510.

\bibitem{Bertram04c}
R.~Bertram, S.~Ganschow, D.~Klimm, P.~Reiche, R.~Uecker, Patent DE 10 2004 003 596 (in German).

\bibitem{Schulz06}
D.~Schulz, S.~Ganschow, D.~Klimm, M.~Neubert, M.~Ro{\ss}berg, M.~Schmidbauer, R.~Fornari, J. Crystal Growth 296 (2006) 27.

\bibitem{Frank65}
F.~C. Frank, Acta Cryst. 18 (1965) 862.

\bibitem{Bates62}
C.~H. Bates, W.~B. White, R.~Roy, Science 21 (1962) 993.

\bibitem{Desgreniers98}
S.~Desgreniers, Phys. Rev. B 58 (1998) 14102.

\bibitem{Logan59}
R.~A. Logan, W.~L. Bond, J. Appl. Phys. 30 (1959) 322.

\bibitem{Jordan80}
A.~S. Jordan, J. Cryst. Growth 49 (1980) 631.

\bibitem{Halvorson72}
J.~J. Halvorson, R.~T. Wimber, J. Appl. Phys. 43 (1972) 2519.

\bibitem{Iwanaga00}
H.~Iwanaga, A.~Kunishige, S.~Takeuchi, J. Mater. Sci. 35 (2000) 2451.

\bibitem{Gadzhiev03}
G.~G. Gadzhiev, High Temperature 41 (2003) 778.

\bibitem{Naumowicz81}
P.~Naumowicz, K.~Wieteska, S.~Szarras, J.~Cinak, Kristall und Technik 16 (1981) 983.

\bibitem{Machida94}
H.~Machida, K.~Hoshikawa, T.~Fukuda, J. Cryst. Growth 137 (1994) 82.

\bibitem{Uecker06b}
R.~Uecker, H.~Wilke, D.~G. Schlom, B.~Velickov, P.~Reiche, A.~Polity, M.~Bernhagen, M.~Rossberg, J. Crystal Growth 295 (2006) 84.

\bibitem{Sato06}
Y.~Sato, H.~Kusumi, H.~Yamaguchi, T.~Komiyama, T.~Aoyama, Physica B 376 (2006) 719.

\bibitem{Sadofev07}
S.~Sadofev, P.~Sch\"afer, Y.-H.~Fan, S.~Blumstengel, F.~Henneberger, D.~Schulz, D.~Klimm, Appl. Phys. Lett. 91 (2007) 201923.

\end{thebibliography}
\end{document}